\documentclass[final,1p,times]{elsarticle} 

\usepackage{graphicx}
\usepackage{amssymb} 
\usepackage{amsthm} 
\usepackage{lineno}

\newcommand{\pt}{$p_{\rm T}$}
\newcommand{\avpt}{$\langle p_{\rm T}\rangle$}
\newcommand{\jpsi}{J/$\psi$}
\newcommand{\psip}{$\psi$(2S)}
\newcommand{\RAA}{$R_{\rm AA}$}
\newcommand{\psipjpsi}{$\psi$(2S)/J/$\psi$}

\journal{Nuclear Physics A} 

\begin{document}

\begin{frontmatter} 

\title{\jpsi\ and \psip\ production in \mbox{Pb-Pb} collisions with the ALICE Muon
Spectrometer at the LHC}

\author{Roberta Arnaldi (for the ALICE\fnref{col1} Collaboration)}
\fntext[col1] {A list of members of the ALICE Collaboration and acknowledgements can be found at the end of this issue.}
\address{INFN Sez. Torino, Via Pietro Giuria 1, I-10126 Torino, Italy}

\begin{abstract} 
Charmonium states are considered important signatures of the strongly interacting
medium created in heavy-ion collisions. 
In the ALICE experiment, these probes can be investigated in the $\mu^{+}\mu^{-}$ 
decay channel, in the forward rapidity region (2.5$<y<$4) down to zero 
transverse momentum.
Results on charmonia production in \mbox{Pb-Pb} collisions at 
$\sqrt{s_{NN}}$=2.76 TeV are presented.   
The centrality and transverse momentum dependence of the inclusive \jpsi\ nuclear 
modification factor are shown and compared with theoretical models. 
Finally, first ALICE results on the \psip\ production in \mbox{Pb-Pb} 
collisions are also discussed.
\end{abstract} 

\end{frontmatter} 


The ALICE experiment at the Large Hadron Collider (LHC) is designed to study 
the formation of the strongly interacting matter in high energy 
heavy-ion collisions. Among the probes of the expected phase transition between 
hadronic and deconfined matter, a relevant role is played by quarkonium states.   
In particular, according to the color-screening model \cite{Sat86}, the in-medium
dissociation probability of such states should provide an estimate of the 
initial temperature reached in the collisions.
Studies performed in the last twenty years at the SPS and RHIC facilities, 
at $\sqrt{s_{NN}}$=17 and 200 GeV respectively, have, indeed, shown a reduction 
of the \jpsi\ production yield beyond the expectations due to the cold nuclear 
matter effects (i.e shadowing and nuclear absorption). 
In spite of the very different center of mass energy, the amount of suppression
observed by SPS and RHIC experiments is similar. 
Furthermore at RHIC a stronger \jpsi\ suppression has been measured at forward 
with respect to mid-rapidity.
These observations suggest the existence of an additional \jpsi\ production
mechanism,  which sets in when higher $\sqrt{s_{NN}}$ are
reached. This mechanism, based on the combination of initially uncorrelated $c$ and 
$\bar{c}$ pairs \cite{Pbm00,And11,The01}, can counteract the quarkonium suppression in the QGP.
Therefore, the measurement of charmonium production is especially promising at 
the LHC, where the high-energy density of the medium and the large number of 
$c\bar{c}$ pairs produced in central Pb-Pb collisions should help to 
disentangle suppression and regeneration scenarios.

The ALICE experiment \cite{Aam08} studies quarkonium production in the
$\mu^{+}\mu^{-}$ decay channel in the rapidity region $2.5<y<4$ and in the 
$e^{+}e^{-}$ decay channel at $|y|<0.9$ \cite{Sco12}. 
In this paper, we present  
quarkonium measurement at forward $y$, while results obtained at mid-rapidity   
are discussed in \cite{Ars12}.

Muons are identified and tracked in a Muon Spectrometer equipped with
a dipole magnet, a set of absorbers, five tracking chambers and a trigger system.
The pixel layers of the Inner Tracking System allow the vertex
determination, while forward detectors such as the VZERO scintillators are used for 
triggering purposes. The VZERO is also used to determine the centrality of the
collisions, through a fit, based on a Glauber model, to its signal 
amplitude \cite{Aam11}.
Results presented in the following are based on a sample of dimuon
triggered events corresponding to an integrated luminosity $L\sim70$ $\mu$b$^{-1}$,
collected during the 2011 \mbox{Pb-Pb} data taking. 
A clean data sample is obtained applying quality cuts to the muon 
tracks. These cuts eliminate mainly tracks hitting the edges 
of the spectrometer's acceptance or crossing the thicker part of the beam 
shield.
Furthermore, to remove hadrons produced in the front absorber, tracks reconstructed in the 
tracking chambers are required to match hits in the trigger planes.

\begin{figure}[htbp]
\begin{center}
\includegraphics[width=0.48\textwidth]{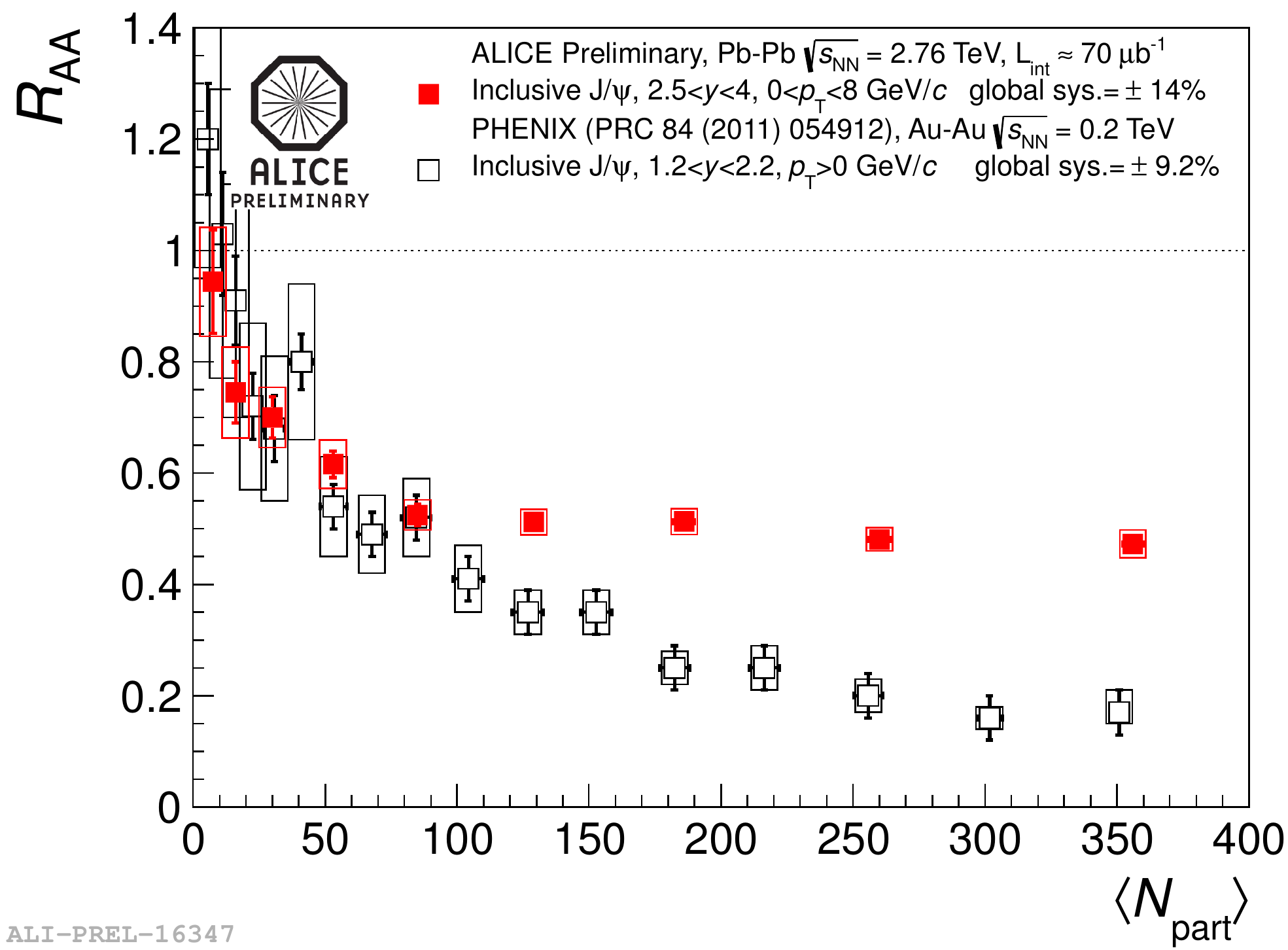}
\includegraphics[width=0.48\textwidth]{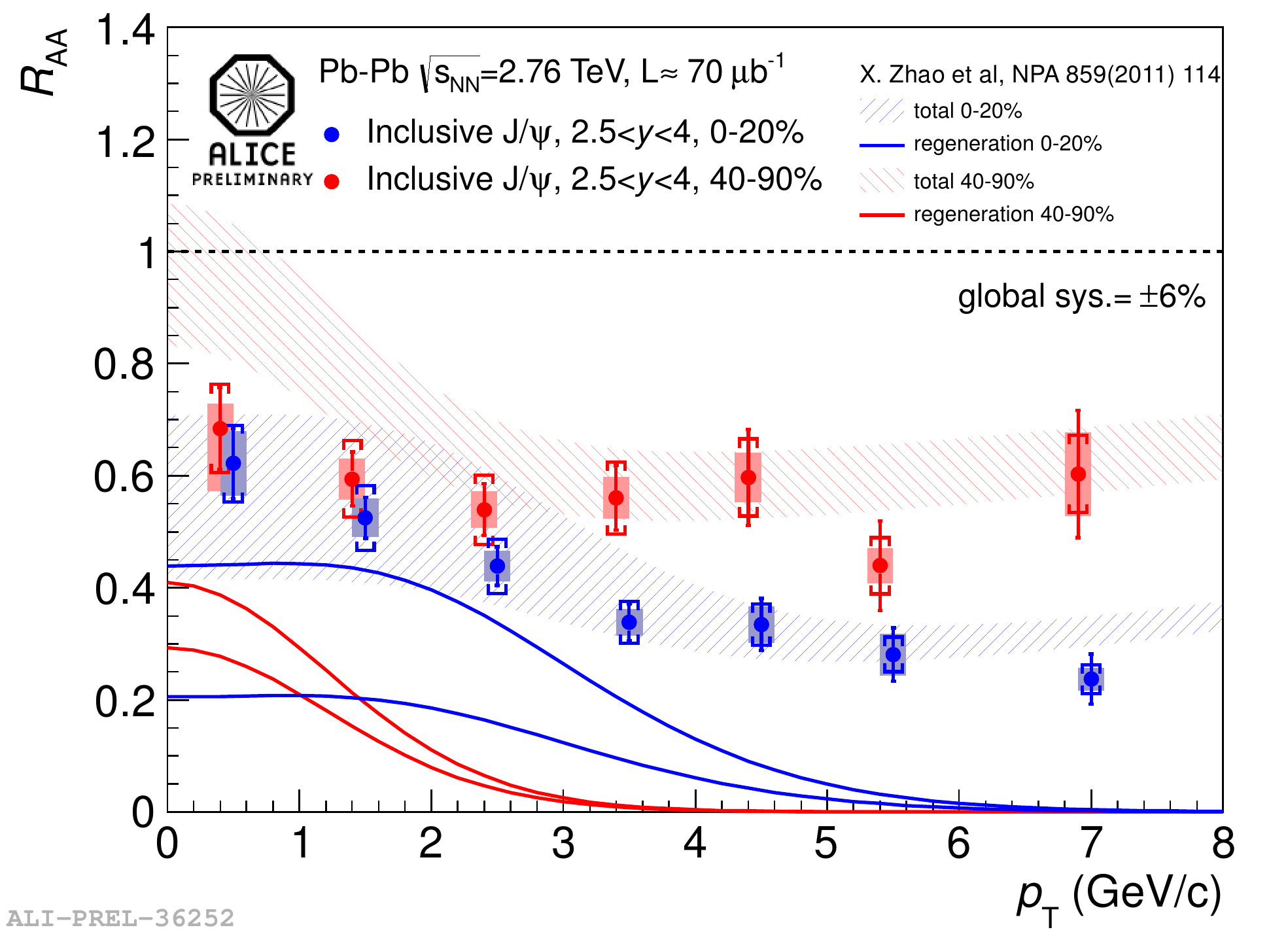}
\end{center}
\caption{Left: the \jpsi\ \RAA\ is shown as a function of $N_{\rm part}$ 
and compared to the PHENIX result. Statistical
errors are shown as lines, while systematic uncertainties are shown as boxes around the
points. Right: the \RAA\ \pt\ dependence is shown for two centrality classes and compared
to a theoretical model \cite{Zha11}. In this case, systematic uncertainties are
split between uncorrelated and partially correlated components, shown,
respectively, as boxes and brackets around the points.
The theoretical bands represent the uncertainty on the 
calculation due to different shadowing assumptions. Global systematic
uncertainties are listed in the legend.
}
\label{fig:fig1}
\end{figure}

The in-medium modification of the \jpsi\ production is quantified through the
nuclear modification factor (\RAA), defined as the ratio of the \jpsi\ yield measured in
\mbox{Pb-Pb} collisions and the expected yield obtained scaling the \mbox{pp} 
\jpsi\ production cross section by the number of binary nucleon-nucleon collisions.
The \jpsi\ yield is extracted by fitting the opposite sign invariant mass
spectrum with an extended Crystal-Ball function (CB2), which allows the
inclusion of non-gaussian
tails on both sides of the \jpsi\ pole. 
The mass position and the width of the CB2 are kept as free parameters in the 
fit, while the tails are tuned on a MC where a \jpsi\ signal is embedded into 
real events.
The background contribution under the \jpsi\ is described by a gaussian 
function with a mass-dependent width. 
Alternatively, an event mixing procedure has been applied, to subtract the 
background contribution, before fitting the signal.
The raw \jpsi\ yield is determined as the average of the results obtained
with the two approaches, including also some modifications of the fitting
procedure (e.g. different CB2 sets of tails or alternative fitting functions).
The corresponding systematic uncertainties on the signal extraction are defined 
as the r.m.s. of these results.
Details on the analysis are given in \cite{Abe12}.  
The total number of \jpsi\ in the kinematic region 0$<p_{\rm T}<$8 GeV/c,
2.5$<y<$4 and in the centrality range 0-90\% amounts to $\sim$40000. 
The high statistics collected in 2011 allows, therefore, a differential study of 
the \jpsi\ production yield, as a function of centrality, $y$ and \pt. 
The procedure described above is
applied in each kinematic bin under study.
To evaluate the \RAA\, the \jpsi\ yield is
divided by the acceptance $\times$ efficiency ($A\times\epsilon$), computed by 
embedding generated \jpsi\ particles into real events. The average $A\times\epsilon$ is
$\sim$14\%, with an 8\% decrease from peripheral to central collisions.
Finally, the \jpsi\ yield measured in \mbox{Pb-Pb} collisions, in each kinematic 
bin, is normalized to the corresponding inclusive \jpsi\ cross section measured in
\mbox{pp} collisions at the same energy \cite{Abe12pp}.
Systematic uncertainties on the \RAA\ depend on the kinematic range 
under study. The main source is due to the \mbox{pp} reference 
($\sim$9$\%$), while other contributions, related, for example, to the choice of the 
MC inputs, to the uncertainty on the trigger, tracking and matching 
efficiency and to the signal extraction amount to less than 6-7$\%$ each.
The inclusive \RAA\, integrated over centrality, \pt\ and $y$ is 
$R_{\rm AA}^{0-90\%}=0.497\pm0.006(stat)\pm0.078(syst)$, exhibiting a clear 
\jpsi\ suppression. 
In our $y$ and \pt\ domain, the contribution from beauty hadron feed-down 
to the inclusive \jpsi\ yield amounts to $\sim$10$\%$, having a negligible effect on 
the \RAA\ measurement.
The centrality dependence of the \RAA, integrated over \pt\ and $y$ is 
shown in Fig.\ref{fig:fig1} (left). 
The pattern observed by ALICE presents a weaker centrality dependence and a smaller
suppression for central collisions with respect to PHENIX results \cite{Ada11},
suggesting a different interplay, at the two energies, of 
suppression and regeneration mechanisms.
Models including a large fraction of \jpsi\ 
produced from pair (re)combination \cite{Zha11,Liu11,Fer12} or all 
\jpsi\ produced at hadronization \cite{And11} can reasonably describe the 
measured \RAA.
The (re)combination contribution is expected to be dominant, especially in 
central collisions, at low \pt, while it becomes 
negligible as the \jpsi\ \pt\ increases.
This behaviour can be investigated by further studying the \RAA\ \pt\ dependence 
in centrality bins, as shown in Fig.\ref{fig:fig1} (right). 
While for the peripheral bin (40-90$\%$) the \pt\ dependence is negligible, in the  most central bin (0-20$\%$) the suppression
increases by $\sim$60$\%$ moving towards low \pt. 
Models including a \pt-dependent contribution from
(re)combination which amounts, at low \pt, to $60\%$ in central and $30\%$ in
peripheral collisions \cite{Zha11,Liu11}, provide a reasonable description of the data.  

\begin{figure}[htbp]
\begin{center}
\includegraphics[width=0.45\textwidth]{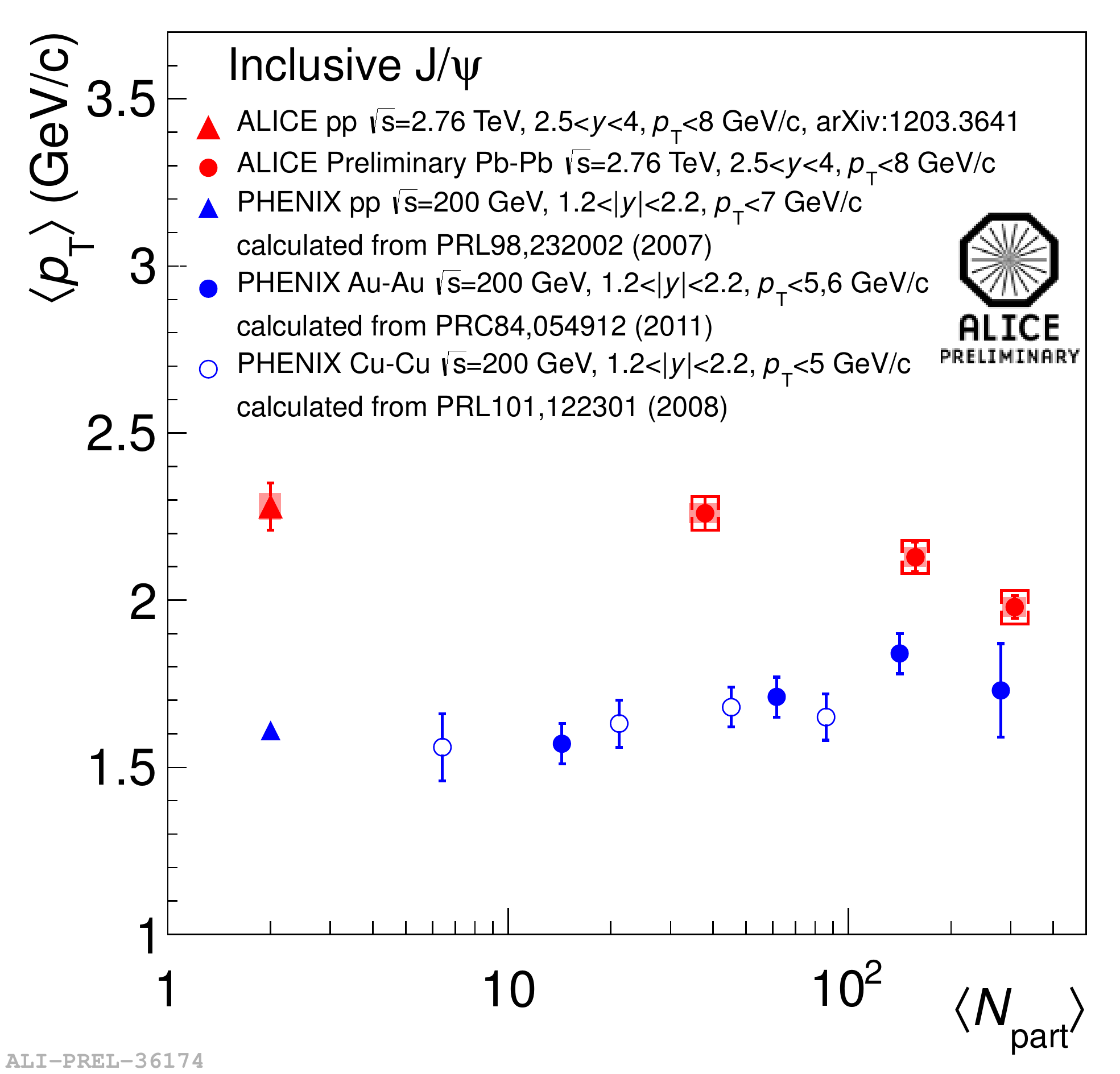}
\includegraphics[width=0.45\textwidth]{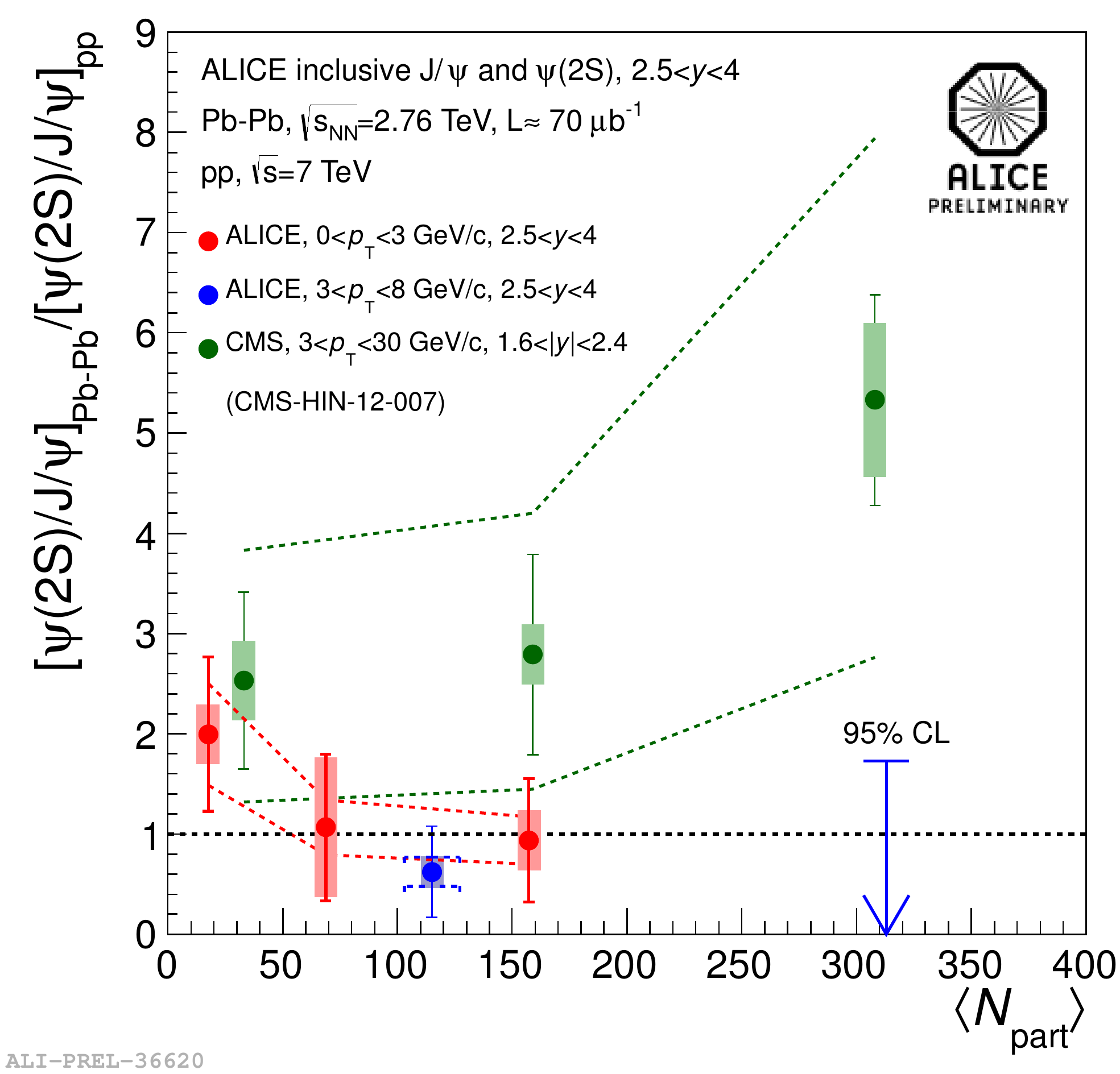}
\end{center}
\caption{Left: the \jpsi\ \avpt\ is shown as a function of $N_{\rm part}$ and is compared
with results from the PHENIX experiment. ALICE statistical errors are shown as lines,
while uncorrelated and partially correlated systematic uncertainties are shown, respectively, as boxes and
brackets around the points.
Right: the double ratio \psipjpsi\ in \mbox{Pb-Pb} and \mbox{pp} is shown versus centrality and 
compared to CMS values \cite{CMS12}. 
Statistical errors are shown as lines, while systematic uncertainties are shown as filled boxes. 
The statistical and systematic uncertainties on the \mbox{pp} reference is shown
as dotted lines.}
\label{fig:fig2}
\end{figure}

Further hints on the \jpsi\ behaviour as a function of \pt\ can be 
inferred from the centrality evolution of the \avpt, extracted from a fit to 
the $d^2N_{J/\psi}/dydp_{\rm T}$ distributions.
As presented in Fig.\ref{fig:fig2} (left), \avpt\ decreases towards central
collisions, pointing to a smaller suppression of low \pt\ \jpsi.
On the contrary, PHENIX results \cite{Ada11,Ada08,Ada07} show a different trend, 
with a \avpt\ increasing with centrality, confirming the different \jpsi\ 
behaviour, versus $\sqrt{s}$, already shown in Fig.\ref{fig:fig1}(left). 

Further insight on charmonium production in \mbox{Pb-Pb} collisions can be
achieved by comparing the \psip\ yield to that of the \jpsi. 
Results are presented as a double ratio of the \psip\ to \jpsi\ yields in \mbox{Pb-Pb} and in 
\mbox{pp} collisions as a function of centrality and in two \pt\ 
classes ($0<p_{\rm T}<3$ GeV/c and $3<p_{\rm T}<8$ GeV/c). 
Signals are extracted with the aforementioned fitting procedure, keeping the \psip\ mass position 
and width fixed relative to the \jpsi\ ones, and then corrected for
the corresponding $A\times\epsilon$. 
For 0-20$\%$ centrality, the very low signal over background ratio (S/B) 
prevents the extraction of the \psip\ yield at low \pt, while in the higher 
\pt\ bin 
(S/B$\sim$0.01) only an upper limit can be evaluated.
Since many of the systematic uncertainties cancel out in the double ratio, the 
main contribution to the systematic error is due the signal extraction, 
being of the order of $\sim$15-60$\%$ depending on the kinematic bin. 
The \mbox{pp} reference has been evaluated at $\sqrt{s}$=7 TeV.
Therefore, we have included a $\sim
$15$\%$ contribution to the systematic uncertainty to take into account a 
possible $\sqrt{s}$-dependence of the
\psipjpsi\ ratio evaluated by comparing 
CDF \cite{Aal09}, LHCb \cite{Aaj12} and CMS \cite{Cha11} results. 
Double ratio results are shown in Fig.\ref{fig:fig2} (right). The large statistics and systematic
uncertainties preclude the drawing of strong conclusions on the \psip\ behavior.  
A significant enhancement in the double ratio for more central collisions 
is not visible in the ALICE data.

We have presented ALICE results on the \jpsi\ \RAA\ as a function of centrality and
\pt\ in \mbox{Pb-Pb} collisions at $\sqrt{s_{NN}}$=2.76 TeV, at forward rapidity.
The \RAA\ shows a clear reduction of the \jpsi\ yield, with a negligible 
centrality dependence and a strong \pt\ dependence, especially in central
collisions.
The \jpsi\ behaviour presents different features with respect to the one 
previously observed by lower energy experiments. 
These features can be qualitatively described by theoretical 
models which include (re)combination as an additional mechanism for \jpsi\ production. 
Further insight in the understanding ot the \jpsi\ behaviour still needs a precise knowledge 
of the cold nuclear matter effects, which will be studied in the incoming \mbox{p-A} data
taking.

\end{document}